\documentstyle[a4wide,epsf,11pt,titlepage]{article}

\newcounter{nref}
\newcommand{\bbib}{%
  \renewcommand{\refname}{\large\bf References}%
  \begin{thebibliography}{99}%
  \setcounter{enumiv}{\thenref}}
\newcommand{\ebib}{%
  \end{thebibliography}%
  \setcounter{nref}{\arabic{enumiv}}}
\newcommand{\head}[3]{%
  \setcounter{nref}{0}%
  \thispagestyle{empty}%
  \section*{\LARGE\bf #1}%
  \stepcounter{section}%
  \addcontentsline{toc}{section}{#1}%
  \large\itshape%
  #2\\\vspace{0.1pt}\\%
  #3%
  \normalsize\upshape%
  \bigskip}

\begin{document}
\head{Neutrino - Photon Interactions at Low Energy }
     {Ara N.~Ioannisian}
     {Yerevan Physics Institute, Yerevan 375036, Armenia  and \\
      Max-Planck-Institut f\"ur Physik (Werner-Heisenberg-Institut), 
F\"ohringer Ring 6, 80805 M\"unchen, Germany}
\subsection*{Abstract}
We discuss $\nu - \gamma$ interactions in the presence
of a homogeneous magnetic field with energies less than pair production
threshold. The neutrinos are taken to be
massless with only standard-model couplings.  The magnetic field
fulfills the dual purpose of inducing an effective neutrino-photon
vertexes and of modifying the photon dispersion relation.
Our conclusion is $\nu - \gamma$ interactions are too weak to be of
importance for pulsar physics.
\vspace{0.5cm}
\subsection*{Introduction}

By now it is well known that in many astrophysical environments the
absorption, emission, or scattering of neutrinos occurs in the presence of
strong magnetic fields ~\cite{r}. Of particular conceptual interest are
$\nu-\gamma$
interactions.
These interactions do not occur in vacuum because neutrinos do not couple
to photons and they are kinematically forbidden.  In the presence of
an external field, neutrinos acquire an effective coupling to photons by
virtue of intermediate charged particles. 
In addition, external fields modify the dispersion relations of all
particles so that phase space is opened for neutrino-photon reactions of
the type $1\to 2+3$.

The processes, which we'll discuss here, are related to the process of
photon splitting that may
occur in magnetic fields ~\cite{adler}.  
In photon splitting the magnetic field plays
the dual role of providing an effective three-photon vertex which does not
exist in vacuum, and of modifying the dispersion relation of the
differently polarized modes such that $\gamma\to\gamma\gamma$ becomes
kinematically allowed for certain polarizations of the initial and final
states.

We would like to stress that $\nu-\gamma$ interactions are important when
typical energies of the particles involving in the reactions are less than
$2\ m_e \simeq 1MeV$. 
If $E > 1MeV$ it becomes important reactions where involve real
electron-positrons and $\nu-\gamma$ interactions become less important. In
the last case one may consider $\nu-\gamma$ processes as radiational
corrections to those where photons do not exist.

Therefore, we are interested to $\nu-\gamma$ processes with typical
energies $E << 1MeV$ ( numerically our results work for $E < 0.5MeV$). 

Energy limitation 
allows us to use the limit of infinitely heavy
gauge bosons and thus an effective four-fermion interaction,
\begin{equation}
{\cal L}_{\rm eff}=
-\frac{G_F}{\sqrt{2}}\,\bar{\nu} \gamma_{\mu}(1-\gamma_5)\nu\,
\bar{E}\gamma^{\mu}(g_V-g_A \gamma_5)E.
\end{equation}
Here, $E$ stands for the electron field,
$\gamma_5=i\gamma^0\gamma^1\gamma^2\gamma^3$,
$g_V=2\sin^2\theta_W+\frac{1}{2}$ and
$g_A=\frac{1}{2}$ for $\nu_e$, and
$g_V=2\sin^2\theta_W-\frac{1}{2}$ and
$g_A=-\frac{1}{2}$ for $\nu_{\mu,\tau}$.
In our subsequent calculations we will always use
$\sin^2\theta_W=\frac{1}{4}$ for the weak mixing angle so that the
vector coupling will identically vanish for $\nu_\mu$ and
$\nu_\tau$.

We'll consider following amplitudes
$\bar\nu \nu$, $\bar\nu \nu \gamma$, $\bar\nu \nu \gamma \gamma$
and related processes.

In the Standard Model neutrino current couple to the electron via vector
or
axial-vector couplings.

For vector coupling, one may get $\nu- \gamma$ interaction amplitudes
using Euler-Heisenberg effective lagrangian of constant electromagnetic
field. 
\begin{eqnarray}
{\cal L}=-{\cal F} -\frac{1}{8\pi^2}\int_0^{\infty}
\frac{ds}{s^3}e^{-m^2_es}[(es)^2L -c.t. ]\\
L=i{\cal G}\frac{\cosh [es\sqrt{2({\cal F}+i{\cal G})}]+
\cosh[es\sqrt{2({\cal F}-i{\cal G})}]}
{\cosh [es\sqrt{2({\cal F}+i{\cal G})}]-
\cosh[es\sqrt{2({\cal F}-i{\cal G})}]}\\
{\cal F}=\frac{1}{4}F_{\mu\nu}^2, 
{\cal G}=\frac{1}{4}F_{\mu\nu}\widetilde{F}_{\mu\nu}, 
\widetilde{F}^{\mu\nu}=\frac{1}{2}\epsilon^{\mu \nu \rho \sigma}F_{\rho\sigma}
\end{eqnarray}

In the Euler-Heisenberg lagrangian we'll assume
\begin{equation}
F_{\mu\nu}=\bar{F_{\mu\nu}}+N_{\mu\nu}+f^1_{\mu\nu}+f^2_{\mu\nu}+...
\end{equation}
here $\bar{F_{\mu\nu}}$ is the external constant electromagnetic field,
$N_{\mu\nu}=
g_v\frac{G_F}{\sqrt{2}}
\{
\partial_{\mu}\bar\nu\gamma_{\nu}(1-\gamma_5)\nu -
\partial_{\nu}\bar\nu\gamma_{\mu}(1-\gamma_5)\nu\}$and $f^i_{\mu\nu}$ are 
the fields of real photons involving in the process. We'll get
$\nu-\gamma$ interaction
amplitudes for vector coupling expanding the Euler-Heisenberg lagrangian
to the Tylor series with respect to the weak fields ($N_{\mu\nu}$,
$f^i_{\mu\nu}$).

Unfortunately we could not use the same procedure for axial-vector
coupling. Therefore we calculated each amplitude separately for
axial-vector coupling.

\subsection*{The $\bf \bar{\nu}\nu$ effective term}

This amplitude will allow us to get dispersion relation of neutrinos in an
external field.
One may expand to the Tylor series the Heisenberg-Euler lagrangian and get
vector part of this
amplitude. We have calculated$^1$ axial-vector part of this amplitude
~\cite{acr}
\begin{eqnarray}
{\it
L}_A&=&\frac{g_AG_F}{\sqrt{2}32\pi^2}\int_0^{\infty}\frac{ds}{s^2}e^{-m^2_es}
[A\partial_{\nu}+B_{\mu\nu}\partial_{\mu}]\bar{\nu}\gamma_{\nu}(1-\gamma_5)\nu\\
A&=&-4e^2s^2{\cal G}\nonumber\\
B_{\mu\nu}&=&\frac{4i}{e^2s^2}\frac{dC_1}{dF_{\mu\lambda}}
\frac{dC_2}{dF_{\lambda\nu}}\\
C_1&=&\frac{2ie^2s^2{\cal G}}{\cosh [es\sqrt{2({\cal F}+i{\cal G})}]-
\cosh[es\sqrt{2({\cal F}-i{\cal G})}]}\\
C_2&=&-2[\cosh [es\sqrt{2({\cal F}+i{\cal G})}]-
\cosh[es\sqrt{2({\cal F}-i{\cal G})}]]
\end{eqnarray}
It is easy to see that the effect is very small for external fields
$E,B \ll \frac{m^2_{W}}{e}$

\subsection*{The $\bar{\nu}\nu \gamma$ effective vertex}

It is shown in ~\cite{ar}
 that the vector part of this matrix element is
proportional to the effective charge radius and may be neglected. Axial
vector part of the amplitude has the form ~\cite{ar}
\begin{equation}
\label{m}
M_5=\frac{g_Ae^3G_F}{\sqrt{2}4\pi^2m_e^2\,e}Z\varepsilon_{\mu}
\bar{\nu}\gamma_{\nu}(1-\gamma_5)\nu\,
\Bigl\{-C_\|\,k_{\|}^{\nu}(\widetilde{F} k)^{\mu}\
+ \ C_\bot\,\Bigl[k_{\bot}^{\nu}(k\widetilde{F})^{\mu}
+k_{\bot}^{\mu}(k\widetilde{F})^{\nu}-
k_{\bot}^2\widetilde{F}^{\mu \nu}\Bigr]\Bigr\},
\end{equation} 
where $\varepsilon$ is the photon
polarization vector. At low energy $C_\bot$ and $C_\|$ are
real functions on $B$.

For $E_{\gamma}<2m_e$ the photon refractive
index always obeys the Cherenkov condition $n>1$ ~\cite{adler}.
Therefore only the Cherenkov process $\nu\to \nu \gamma$ is kinematically 
allowed. 

The four-momenta conservation constrains the photon emission angle to have
the value
\begin{equation}\label{emissionangle}
\cos \theta = \frac{1}{n} \
\left[1+(n^2-1)\frac{\omega}{2E}\right],
\end{equation}
where $\theta$ is the angle between the emitted photon and incoming
neutrino.
It turns out that for all situations of practical interest we have
$|n-1|\ll 1$ ~\cite{adler}.
This reveals that the outgoing photon
propagates parallel to the original neutrino direction.

It is interesting to compare this finding with the standard 
plasma decay process $\gamma\to\bar\nu\nu$ which is dominated by the
$\Pi^{\mu \nu}$. Therefore, in the approximation
$\sin^2\theta_W=\frac{1}{4}$ only the electron
flavor contributes to plasmon decay. Here the Cherenkov rate is equal for 
(anti)neutrinos of all flavors.

For neutrinos which propagate perpendicular to the magnetic
field, a Cherenkov emission rate can be written in the form
\begin{eqnarray}\label{finalresult}
\Gamma\ \approx  = \ 
2.0\times10^{-9}~{\rm s}^{-1}~\left(\frac{E_{\nu}}{2m_e}\right)^5
\left(\frac{eB}{m^2_e}\right)^2 h(B),
\end{eqnarray}
where 
\begin{equation}
h(B)= 
\cases{(4/25)\,(eB/m^2_e)^4&for $eB\ll m^2_e$,\cr
1&for $eB\gg m^2_e$.\cr}
\end{equation}
Turning next to the case $E>2m_e$ we note that in the presence of a
magnetic field the electron and positron wavefunctions are Landau
states so that the process $\nu\to\nu e^+e^-$ becomes kinematically
allowed. Therefore, neutrinos with such large energies will
lose energy primarily by pair production rather than by Cherenkov
radiation ~\cite{ara10}.

\subsection*{The $\bar{\nu}\nu \gamma \gamma$effective vertex}

In the vacuum this amplitude is highly suppressed. The axial-vector
coupling is zero due to Landau-Yang theorem ( two photons cannot have
total spin equal to one) in
four-Fermi limit.
Beyond four-Fermi limit this amplitude is double Fermi suppressed. 
In ~\cite{dr1} it was found that $\bar{\nu}\nu \to \gamma \gamma$ (or
crossed versions) cross-section is
$\sigma \sim 10^{-68}(\frac{\omega}{MeV})^6 cm^2$.

However, this amplitude is not double Fermi suppressed in a magnetic field.
In ~\cite{s} it was shown that in a weak magnetic field 
($B \ll \frac{m_e^2}{e}$)
vector coupling is dominant and cross section of the process is
$\sigma \sim 10^{-50}(\frac{\omega}{MeV})^6 (\frac{eB}{m_e^2})^2cm^2$.

In a strong magnetic field \cite{ak} contribute both- vector and axial
vector couplings. Since $B \gg \frac{m_e^2}{e}$ in the amplitude survive
only lowest Landau levels and we lost dependence on magnetic field 
$\sigma \sim 10^{-50}(\frac{\omega}{MeV})^6cm^2$.

\subsection*{The $\bar{\nu}\nu \gamma\gamma\gamma$ effective vertex}

In this amplitude contribute only vector coupling ~\cite{dr}. The cross
section,~\cite{dr}
of
the process $ \bar{\nu}\nu \to \gamma\gamma\gamma$ (or crossed processes
$\nu \gamma \to \nu \gamma \gamma$ and 
$\gamma\gamma \to \bar{\nu}\nu\gamma$), is 
$\sigma \sim 10^{-50}(\frac{\omega}{MeV})^{10} cm^2$.

\subsection*{Conclusions}

The
strongest magnetic fields known in nature are near pulsars. However,
they have a spatial extent of only tens of kilometers. Therefore, even
if the field strength is as large as the critical one, most neutrinos
escaping from the pulsar or passing through its magnetosphere will not
interact with photons. Thus, the magnetosphere of a pulsar is quite
transparent to neutrinos.
Therefore our main conclusion is $\nu-\gamma$ interactions are too weak to
be of
practical importance for pulsar physics.

\subsection*{Acknowledgments}

It is pleasure to thanks the organizers of the Neutrino Workshop at the
Ringberg Castle for organizing a very interesting workshop. A.I.
acknowledges support from the A.von Humboldt Foundation.


\begin{thebibliography}{9}
\bibitem{r} G. G. Raffelt, {\it Stars as Laboratories for  
  Fundamental Physics\/} (University of Chicago Press, Chicago, 1996).
\bibitem{adler} S. L. Adler, {\it Ann. Phys.} (N.Y.) {\bf 67}(1971) 599.
\bibitem{acr} A. N. Ioannisian, G. Raffelt and C. Schubert, work in
progress.
\bibitem{ar} A. Ioannisian  and G. Raffelt, {\it Phys. Rev.}  {\bf D55}
(1997) 7038.
\bibitem{ara10} A. V. Borisov, A. I. Ternov and V. Ch. Zhukovsky,
  {\it Phys. Lett.} {\bf B318} (1993) 489.
  A. V. Kuznetsov and N. V. Mikheev, 
  {\it Phys. Lett.} {\bf B394} (1997) 123. 
\bibitem{dr1} D. A. Dicus and W. W. Repko, {\it Phys. Rev.}  {\bf D48}
(1993) 5106.
\bibitem{s} R. Shaisultanov, {\it Phys. Rev. Lett.} {\bf 80} (1998) 1586.
\bibitem{ak} A. N. Ioannisian and N. A. Kazarian, work in
progress.
\bibitem{dr} D. A. Dicus and W. W. Repko, {\it Phys. Rev. Lett.} {\bf 79}
(1997) 569.
\end{thebibliography}
\end{document}